\documentclass[preprint,12pt]{elsarticle}

\usepackage{graphicx}

\usepackage{amssymb}

\usepackage{amsmath}
\usepackage{subfig}
\usepackage{multirow}
\usepackage{threeparttable}
\usepackage{booktabs}

\journal{Physica A}

\begin{document}

\begin{frontmatter}



\title{The scaling of human mobility by taxis is exponential}


\author[bhu]{Xiao Liang}
\ead{liangxiao@nlsde.buaa.edu.cn}

\author[bhu]{Xudong Zheng}

\author[bhu]{Weifeng Lv}

\author[bhu]{Tongyu Zhu}

\author[bhu]{Ke Xu\corref{cor1}}
\ead{kexu@nlsde.buaa.edu.cn}

\address[bhu]{State Key Laboratory of Software Development Environment, Beihang University, Beijing 100191, People's Republic of China}
\cortext[cor1]{Corresponding author}

\begin{abstract}
As a significant factor in urban planning, traffic forecasting and prediction of epidemics,
modeling patterns of human mobility draws intensive attention from researchers for decades.
Power-law distribution and its variations are observed from quite a few real-world human mobility datasets
such as the movements of banking notes, trackings of cell phone users'
locations and trajectories of vehicles. In this paper, we build models for 20 million trajectories with
fine granularity collected from more than 10 thousand taxis in Beijing. In contrast to most
models observed in human mobility data, the taxis' traveling
displacements in urban areas tend to follow an
exponential distribution instead of a power-law. Similarly, the elapsed time can
also be well approximated by an exponential distribution. Worth mentioning, analysis of the
interevent time indicates the bursty nature of human mobility, similar to many other human activities.

\end{abstract}


\end{frontmatter}


\section{Introduction}
The emergence of location tracking devices (e.g., GPS navigator
and mobile devices), and location-based services (LBS, e.g., Foursquare,
Yelp check-in and Google places) provide unprecedented opportunities
to study human mobility patterns from trillions of trails and footprints,
which are of great significance in urban planning \cite{Rozenfeld2008b}, traffic forecasting \cite{Jiang2009},
marketing campaign \cite{Agliari2010}, prediction of epidemics \cite{Hufnagel2004}
and designing of mobile network protocols \cite{Lee2009}.

Researchers observe that people's daily activities (e.g., commuting between
home and workplace, going shopping, and going vacation) follow reproducible mobility patterns
\cite{Gonzalez2008,Song2010,Choujaa2010} by studying trackings of cell phone users' locations.
Similar results are also discovered from GPS trackings \cite{Zheng2008,Jiang2009,Rhee2008,Bazzani2010,Jiang2011a,Veloso2011},
wireless network traces \cite{Kim2006}, check-ins from location-based services
\cite{Cheng2011,Cho2011} and even movements of banking notes \cite{Brockmann2006}.

Be specific, Brockmann et al. \cite{Brockmann2006} observe that the distribution of
displacements between consecutive reported sightings of the banking notes 
follows a power-law, and they conclude that human travel behavior can be
described by L\'{e}vy walks with heavy-tailed pause time. Similarly, \cite{Rhee2008}
and \cite{Jiang2009} observe that human trajectories data
from GPS trackings could be approximated by L\'{e}vy flights. And researchers
even notice obvious L\'{e}vy flights based on mobility patterns from
trails of animals \cite{Viswanathan1996,Atkinson2002}. Despite the
prevailing L\'{e}vy walks, González et al. \cite{Gonzalez2008}
discover existence of spatio-temporal regularity in human movements,
indicating people are very likely to return to a few frequently
visited locations. A high confidence in predicting human movements is found
due to the underlying reproducibility of human mobility patterns by Song et al
\cite{Song2010}. The authors also propose a new model to characterize human mobility
patterns \cite{Song2010a}. Han et al. \cite{Han2011} demonstrate that the scaling
law in human mobility could be explained by hierarchy of the traffic systems.

However, the datasets mentioned above also have their limitations. For example, banking notes
perhaps are deposited in banks and then withdrawn after a long time, or transferred
for many times between consecutive observations. As for trackings of cell-phone users,
when people initiate or receive calls and text messages, they are probably on the way
from their origins to destinations. Furthermore, the granularity of some datasets are
coarse, and the trajectories are between cities that are far from each other, and the 
locations recorded may have large deviations from people's actual movements.

In this paper, we study human mobility patterns for 20 millions of trajectories extracted
from more than 10,000 taxis in urban areas of Beijing. Comparing to the other datasets mentioned
above, the granularity of the taxis trails are in very fine granularity. Besides, the data
might also reveal the effects of the urban traffic network on people's movements.

The rest of the paper is organized as follows. In Section \ref{sec:pre} we introduce the
method of model selection. Section \ref{sec:datades} describes the data used in the paper.
We show our analysis and discuss the findings in Section \ref{sec:ana}. Finally, we conclude
in Section \ref{sec:con}.

\section{\label{sec:pre} Preliminary} 
Model selection is to identify
the most appropriate model that is supported by the actual data from
candidate models. Contrary to maximizing fit and null hypothesis
tests, model selection criteria consider both fits with the data and
complexity of models, and enable to compare multiple models at the
same time. There are two criteria commonly used, which are Akaike
information criterion (AIC) and Bayesian information criterion (BIC) \cite{Johnson2004,Hastie2008}.

In this paper, we mainly employ AIC to compare two models: a power-law
$y=Ax^{-\alpha}$ and an exponential $y=Be^{-\lambda x}$, 
where $A$ and $B$ are normalization constants. The steps of model selection are
shown as follows.

\begin{enumerate}
\item \label{estimate} Estimating the parameters of models using
  maximum likelihood method. The details about how to perform maximum
  likelihood estimates (MLE) of these models can
  refer to \cite{Clauset2009}, \cite{Edwards2007} and \cite{Mashanova2010}.
\item Calculating the AIC scores for the models. The AIC score for
  model $i (i \in \{1,2\})$ is given by
$$AIC_i = -2\log L_i+ 2K_i$$
Here $L_i$ is the likelihood in which parameters are assigned with the
estimated values from step \ref{estimate} and $K_i$ is the number of
parameters in the model $i$.
\item Determining the best models. The Akaike weights can be
  considered as relative likelihoods being the best model for the
  observed data. Let
\begin{eqnarray*}
AIC_{min}=\min_{i\in\{1,2\}}{\{AIC_i\}}\\
\Delta_i = AIC_i-AIC_{min} , i\in \{1,2\}
\end{eqnarray*}
Then, the Akaike weights are represented by
$$W_i=\frac{e^{-\Delta_i/2}}{\sum_{j=1}^{2}{e^{-\Delta_j/2}}}  ,i\in
\{1,2\}$$ So the model corresponding with the largest Akaike weight
should be selected as the best one.
\end{enumerate}

\section{\label{sec:datades}Data description}
To explore the urban movements, we use two GPS data sets, which were
generated by over 10,000 taxies in Beijing, China, from Oct. 1st, 2010
to Dec. 31st, 2010($D_1$) and from Oct. 1st, 2008 to Nov. 30th,
2008($D_2$) respectively. The GPS data from every taxi were collected
at about 1-minute intervals. In addition to location $l$ (longitude,
latitude) and instantaneous velocity $v$, operational status $s$ (with
passengers or without passengers) of taxis was also captured from
GPS equipment at the same time. Thus, the GPS data for a taxi $k$
at time $t$ can be denoted by a tuple $(k, l, v, s, t)$. When a
taxi was carrying customers, it offered the proxy to understand
human mobility patterns in urban areas.

From both data sets, tens of millions of human trajectories can be
extracted. More specifically, according to operational status
collected, we can identify when and where customers got into and got
off the same taxi. Therefore, a trajectory can be uniquely represented
by $(k, l_O, t_O, l_D, t_D)$, which means that the customers departed
from the origin $l_O$ at the time $t_O$ to the destination $l_D$ at
the time $t_D$ with the aid of the taxi $k$. The tuples with the time
$t$ between $t_O$ and $t_D$ for the taxi $k$ were associated with the
trajectory. Here the displacement $\Delta L$ and the elapsed time
$\Delta T$ for the trajectory can be calculated as follows:
\begin{eqnarray*}
\Delta l & = & | l_D - l_O |\\
\Delta T & = & | t_D -  t_O |
\end{eqnarray*}
If the next trajectory of the taxi $k$ is $(k, \tilde l_O,
\tilde t_O, \tilde l_D, \tilde t_D)$, the interevent time $\tau$ that
is the duration without passengers can be computed by \[\tau = |\tilde
t_O-t_D|\]

However, there were some abnormal tracks that need to be excluded
from the results. For example, when $\Delta T < 1 \text{ min}$ and $\Delta T > 120 \text{ min}$, 
the trajectory should be considered invalid because passengers seldom readily spend too short or 
too long time on taking taxis. Finally, we
derive 12,028,929 trajectories in $D_1$ and 9,942,697 ones in $D_2$ as
shown in Table \ref{tab:datasets}. It is worthy to note that our analyses are mainly based
on $D_1$ whereas the $D_2$ intends to be compared with $D_1$ to detect
some changes of trends.

\begin{table}[htbp]
  \caption{The numbers of trajectories and taxis in both datasets}
  \label{tab:datasets} 
  \centering 
  \begin{tabular}{cccc}
    \hline
   {} & {} & Number of trajectories & Number of taxis\\
    \hline
    \multirow{3}{*}{$D_1$} & Oct. 2010 & 3,932,107 & 11,686\\
    {} & Nov. 2010 & 3,754,405 & 11,636\\
    {} & Dec. 2010 & 4,342,417 & 11,562\\
    {} & Total & 12,028,929 & 11,686\\
    \hline
    \multirow{2}{*}{$D_2$} & Oct. 2008 & 5,111,144 & 10,695\\
    {} & Nov. 2008 & 4,831,553 & 10,671\\
    {} & Total & 9,942,697 & 10,695\\
    \hline
 \end{tabular} 
\end{table}

\section{\label{sec:ana}Statistical results and explanations}
In this section, we first investigate the displacements between
origins and destinations. Then the elapsed time of travels is
analyzed and the correlation between displacement and elapsed time is revealed. 
Finally, the duration with no passengers (i.e., interevent time) is studied, which
suggests human travel demands.

\subsection{\label{subsec:disp}Displacement}  For trajectories, the
pairs of origin and destination reflect the purposes of human
movements directly. As we know, a displacement is the distance of a line segment connecting the origin and
destination. Contrary to the actual path traversed
from the origin to destination in the city, the displacement is not
influenced by geographic constraints and artificial interference. It is better to
characterize the human mobility in urban areas.

\begin{figure}[htbp]
  \centering \subfloat[]{
    \label{fig:disp_a}
    \includegraphics[scale=.3]{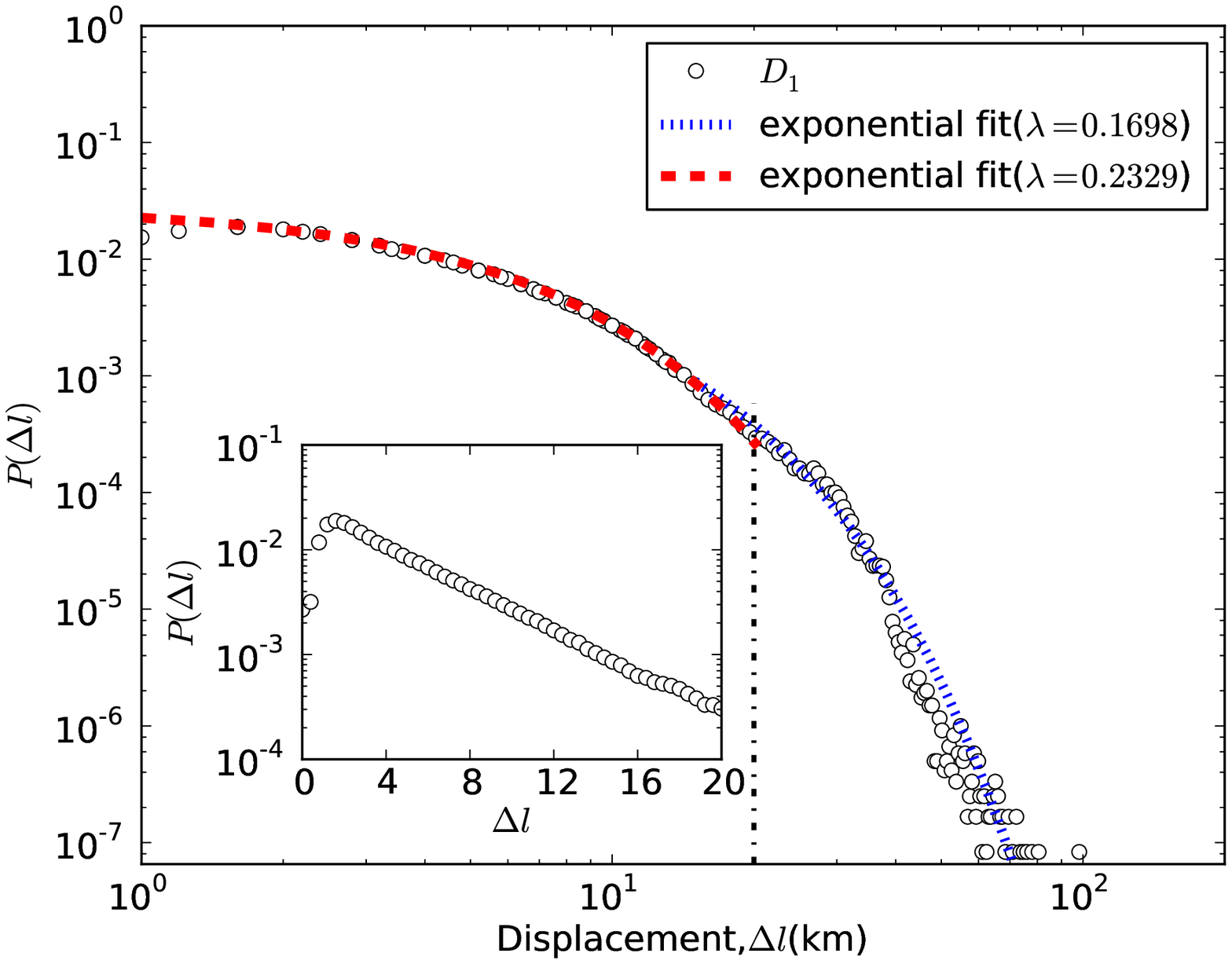}
  } \subfloat[]{
    \label{fig:disp_b}
    \includegraphics[scale=.3]{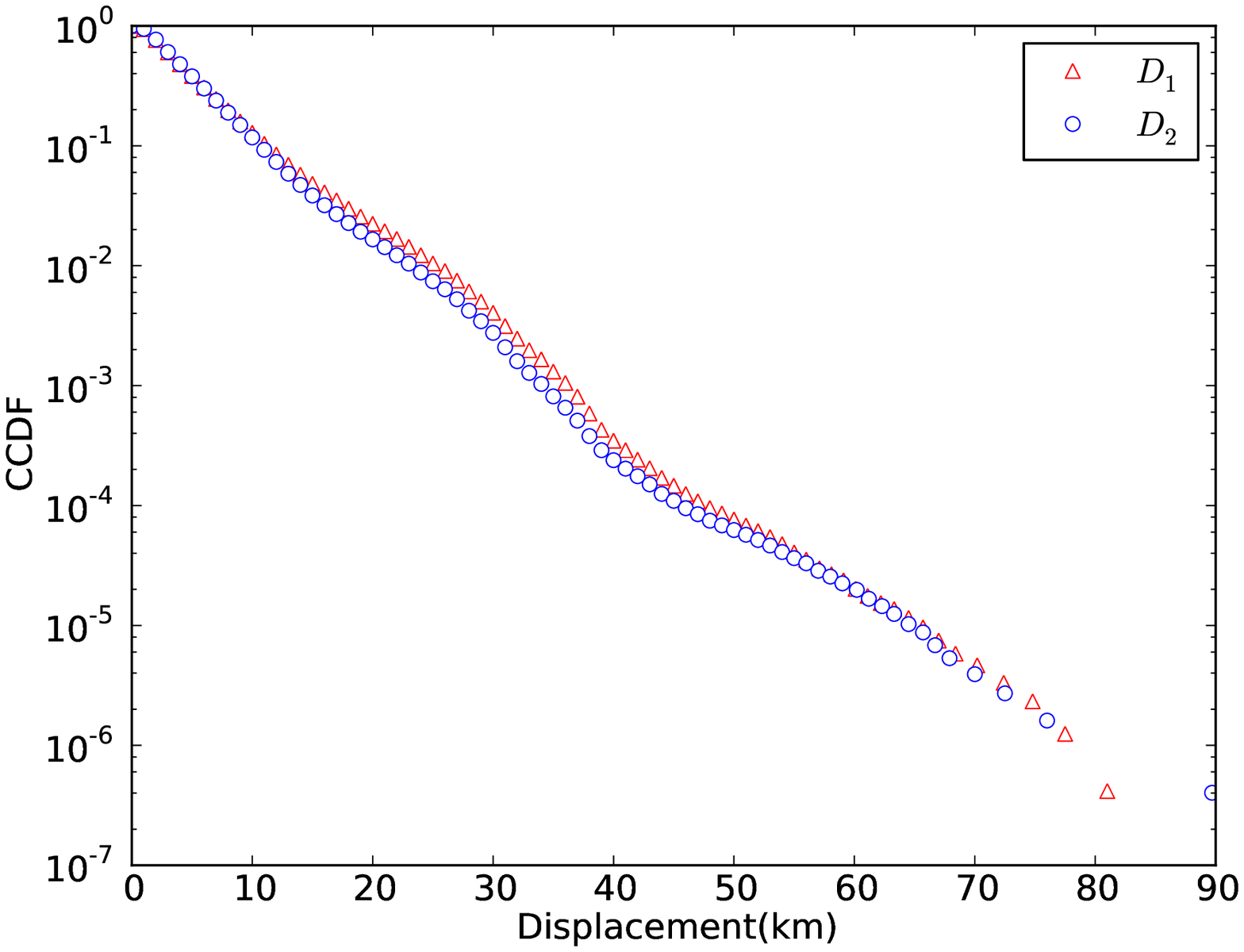}
  } \caption{ Displacements of trajectories. (a) The probability
    density function $P(\Delta l)$ obtained for studied dataset
    $D_1$. The inset shows the first part of $P(\Delta l)$ in semi-log
    scale. The red dashed line indicates an exponential with measured exponent
    $\lambda=0.2329$. The blue dotted line represents an exponential
    with measured exponent $\lambda = 0.1698$. (b) The CCDFs of
    displacements for both datasets.}
\end{figure}

Therefore, we measure the probability $P(\Delta l)$ with a
displacement $\Delta l$ in the dataset $D_1$. As
shown in Figure \ref{fig:disp_a}, $P(\Delta l)$ increases suddenly in
the beginning and reaches the peak when $\Delta l$ is about 2 km. After
that, it decreases dramatically. The reason resulting in the rise at
first is that the distances of human travels
are seldom very short. Moreover, there are approximately
98\% trajectories traversing a distance of less than 20 km, also
suggesting that most movements occurred in urban areas. Intuitively, most individuals seem to be more
apt to wander in the neighborhood of some places (e.g., homes, schools,
and workplaces) in daily life. Consequently, the statistical distribution
agrees with our experiences well.

Given the distribution of displacements,  we
partition it into two parts according to the displacement of
20km. For the first part, most displacements occurred in urban
areas, while for the latter part, these displacements often occurred
between urban areas and suburbs. Because of economic consideration,
few people choose taxis to traverse a large displacement. It seems
that the distribution shows different trends in the two parts.
Therefore, we utilize the AIC mentioned in Section \ref{sec:pre}
to compare two frequently used models: a power-law $P(\Delta l) \sim
{\Delta l}^{-\alpha}$ and an exponential $P(\Delta l) \sim e^{-\lambda
  \Delta l}$ for the two parts. The detailed results for model selection are illustrated
in Table \ref{tab:modelsel}. From the table, we can conclude the distribution of
displacements can be well fitted by an exponential distribution with
an exponential cutoff because of $W_{exp}\gg W_{pow}$ in the two
parts. Moreover, the two different exponential fits
observed in Figure \ref{fig:disp_a} also support the conclusion
further. We also notice that the exponential exponent for the first part is
larger than the one for the latter part. It is reasonable because people
travelling large displacements by taxis are less sensitive to taxi fares,
resulting in decreasing slightly slower in the tail of distribution. 
In addition, it is interesting that the power-law exponents
obtained for the first part in both datasets are not far from ones
observed in \cite{Brockmann2006} ($\mu =
1.59$) and \cite{Gonzalez2008} ($\mu = 1.75$). 
From above results, it can be inferred that displacements occurred in urban
areas are distributed according an exponential rather than a power-law.

Furthermore, the distributions of displacements for both datasets are
plotted respectively in Figure \ref{fig:disp_b}. In order to reduce
errors in the tail of distribution, we use complementary cumulative
distribution function (CCDF) here. It appears that both distributions are
almost coincident. Besides, the
results of MLE for two data sets are very close to each other as shown
in Table \ref{tab:modelsel}. These all indicate that human travel patterns have no
obvious changes in recent years.

\begin{table}[htbp]
\begin{threeparttable}
  \caption{Results of model selection for displacement in both
    datasets.}
  \label{tab:modelsel} 
  \centering 
  \begin{tabular}{ccccc}
   \hline
   {Part} & \multicolumn{2}{c}{Model}  & $D_1$ & $D_2$\\
   \hline
   \multirow{8}{*}{First part} & \multirow{4}{*}{Power law} & {MLE for $\alpha$} & {1.4869} & {1.5108}\\
   {} & {} & (95\% CI)\tnote{\S} & (1.4859, 1.4880) & (1.5097, 1.5119)\\
   {} & {} & $R^2$\ \tnote{\dag} & 0.9112 & 0.8950\\
   {} & {} & $W_{pow}$\tnote{\ddag} & 0 & 0\\
   \cline{2-5}
   {} & \multirow{4}{*}{Exponential} & {MLE for $\lambda$} & \textbf{0.2329} & \textbf{0.2403}\\
   {} & {} & (95\% CI) & (0.2328, 0.2331) & (0.2401, 0.2405)\\
   {} & {} & $R^2$ & 0.9989 & 0.9995\\
   {} & {} & $W_{exp}$\tnote{\ddag} & 1 & 1\\
   \hline
  \multirow{8}{*}{Latter part} & \multirow{4}{*}{Power law} & {MLE for $\alpha$} & {5.1048} & {5.2729}\\
   {} & {} & (95\% CI) & (5.0888, 5.1208) & (5.2519, 5.2939)\\
   {} & {} & $R^2$ & 0.9119 & 0.9165\\
   {} & {} & $W_{pow}$ & 0 & 0\\
   \cline{2-5}
   {} & \multirow{4}{*}{Exponential} & {MLE for $\lambda$} & \textbf{0.1698} & \textbf{0.1768}\\
   {} & {} & (95\% CI) & (0.1692, 0.1705) & (0.1760, 0.1777)\\
   {} & {} & $R^2$ & 0.9707 & 0.9732\\
   {} & {} & $W_{exp}$ & 1 & 1\\
   \hline
 \end{tabular} 
 \begin{tablenotes}
   \item[\S] A 95\% confidence interval.
   \item[\dag] Coefficient of determination to measure the goodness of fit of a model.
   \item[\ddag] Akaike weights representing relative likelihoods of models.
 \end{tablenotes}
\end{threeparttable}
\end{table}

However, the observed shape of $P(\Delta l)$ may be caused by
geographic heterogeneity (i.e., there are different statistical
properties of human travels among distinct locations). We will intend
to discuss the influences of geography on human mobility below. Here
five representative local areas of circular regions with radius
1 km are selected, including Beihang university (BHU), Beijing Railway
Station(BRS), Beijing Capital International Airport (BCIA),  Xidan(a
business district) and a residential district. These areas often have
distinct transport features and population density. It is appropriate
to exploit them to investigate the discrepancies caused by geographic
heterogeneity.

\begin{figure}[htbp]
  \centering
  \includegraphics[scale=.4]{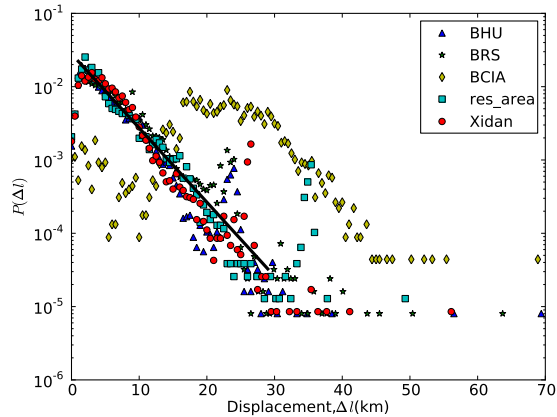}
  \caption{Distributions of displacements in different areas. The
    black line represents the exponential fit with exponent 0.2329,
    which applies to overall distribution for $D_1$.}
  \label{fig:disp_allpos}
\end{figure}

The five distributions of displacements initiated in these regions for
the $D_1$ dataset are plotted in Figure \ref{fig:disp_allpos}
separately. The obtained distributions, except BCIA, agree with each
other very
well and can be well approximated by the exponential fit, which
also applies to the overall distribution of displacements as referred
above. Also, most travels occurred in these areas have short
displacements of less than 20 km. Nevertheless, movements initiated in
BCIA often have larger displacements on average and show evident
differences with other areas. The reason is that BCIA is located
suburb of Beijing. Therefore, passengers who got off planes usually
have to go to urban areas traversing long distances by taxis. At the
same time, the tail of distribution appears to decay exponentially in
the same trend with the others. In this paper we mainly focus on the
movements in urban areas, so it can be concluded that the geographic
heterogeneity does not affect our results on the observed human
mobility patterns. The intrinsic travel demands in urban environments
lead to the statistical pattern.

It must be noted that our findings are not coincident with those obtained
from bill notes \cite{Brockmann2006} and mobile
phones \cite{Gonzalez2008} which have demonstrated that the
distribution of displacements could be well approximated by a
power-law or a truncated power-law. According to the characteristics of
GPS data from taxis, there are probably two reasons that can be used to
explain the inconsistency. Firstly, the movements involved in the
datasets occurred mostly in urban areas and had small
displacements. However, in other datasets, people travelled around the
whole country by different transportation means and could traverse a
long distance of more than 1000 km. The travels with long distances
often happened between cities. So the displacements in urban
environments decay faster than those measured from other
datasets. Secondly, taxi fare increases with the distance passed
by. When people intend to travel long distances, they may choose
other public transportation means rather than taxis out of economic
consideration. As a result, trajectories with long displacements may
have lower likelihood to be captured by taxis. In summary, because of
these two factors, the probability distribution of displacements in
urban areas decays more quickly than those in other datasets and
is inclined to be exponential. To what degree the two
factors affect human travel demands in cities should be
studied on more datasets further.

In addition, \cite{Jiang2009} study the
travels of only 50 taxis collected from four cities in Sweden. They
find the distribution of trail length follows a double power-law,
implying both intracity and intercity movements each show a scale
property. In their results, we notice that distribution in cities still follows
a power-law in spite of economic effects. Compared with their dataset, our
datasets are more comprehensive and cover the whole city
fully. Also, through analyzing GPS traces of taxis in Lisbon,
\cite{Veloso2011} illustrate that trip distance can be well represented
by an exponential distribution. It is worthy to note that
the exponential parameter they obtained ($\lambda = 0.26$) is not far from
ours. \cite{Bazzani2010} consider GPS data of private cars in
Florence urban areas. It is observed that total distance of daily round
trips for each vehicle agrees with an exponential distribution very well. They also
identify vehicle stop positions and discover single trip length distribution 
deviates from an exponential and favors a power law gradually when trip length becomes longer.
In summary, the results of urban mobility in \cite{Veloso2011} and
\cite{Bazzani2010} are consistent with ours. 
Thus, it can be conjectured that the phenomenon may not happen
accidentally and exist in urban areas of cities widely.

\subsection{Elapsed time}
\begin{figure}[htbp]
  \centering \subfloat[]{
    \label{fig:elapse_a}
    \includegraphics[scale=.3]{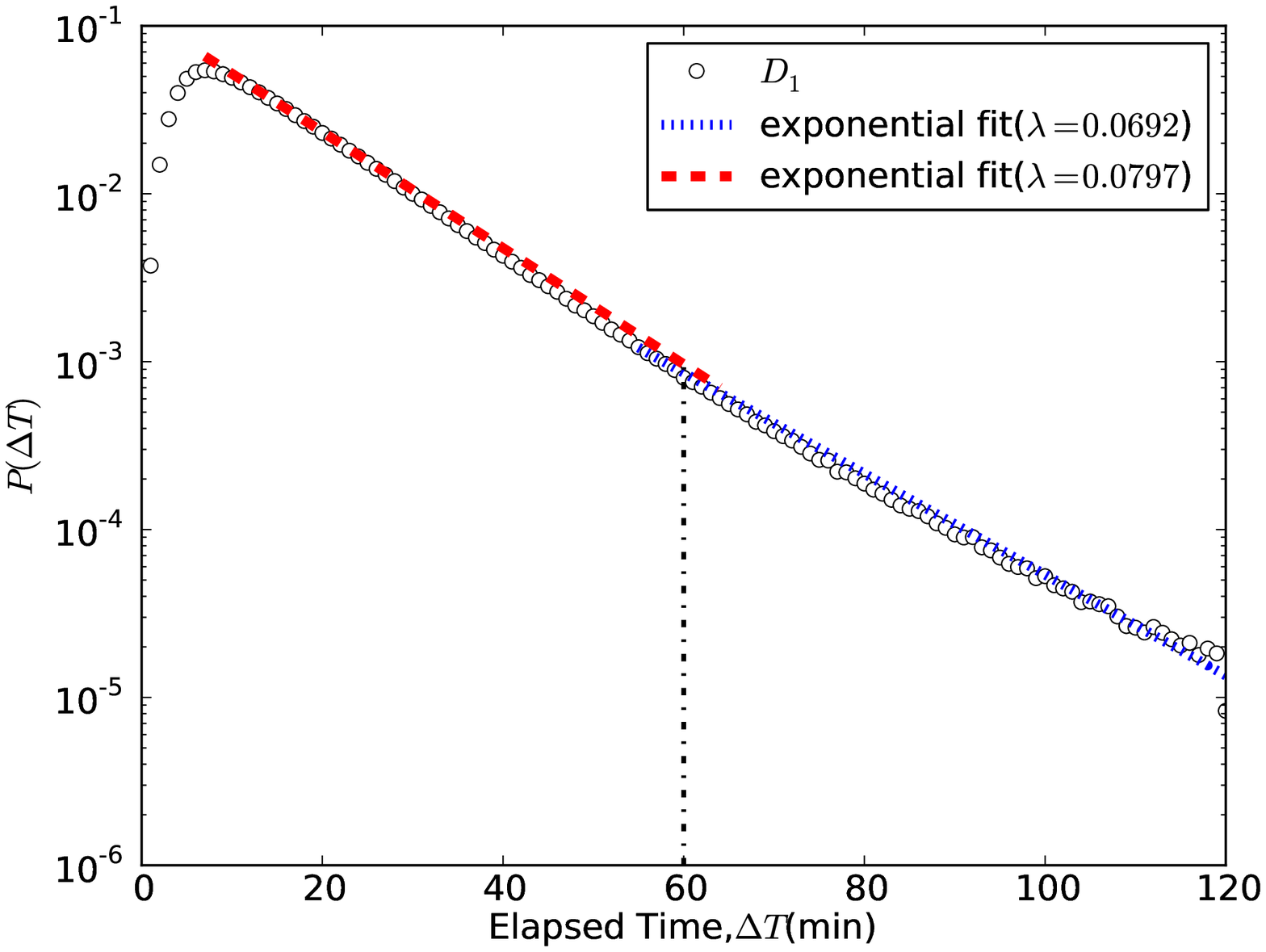}
  }  \subfloat[]{
    \label{fig:elapse_b}
    \includegraphics[scale=.3]{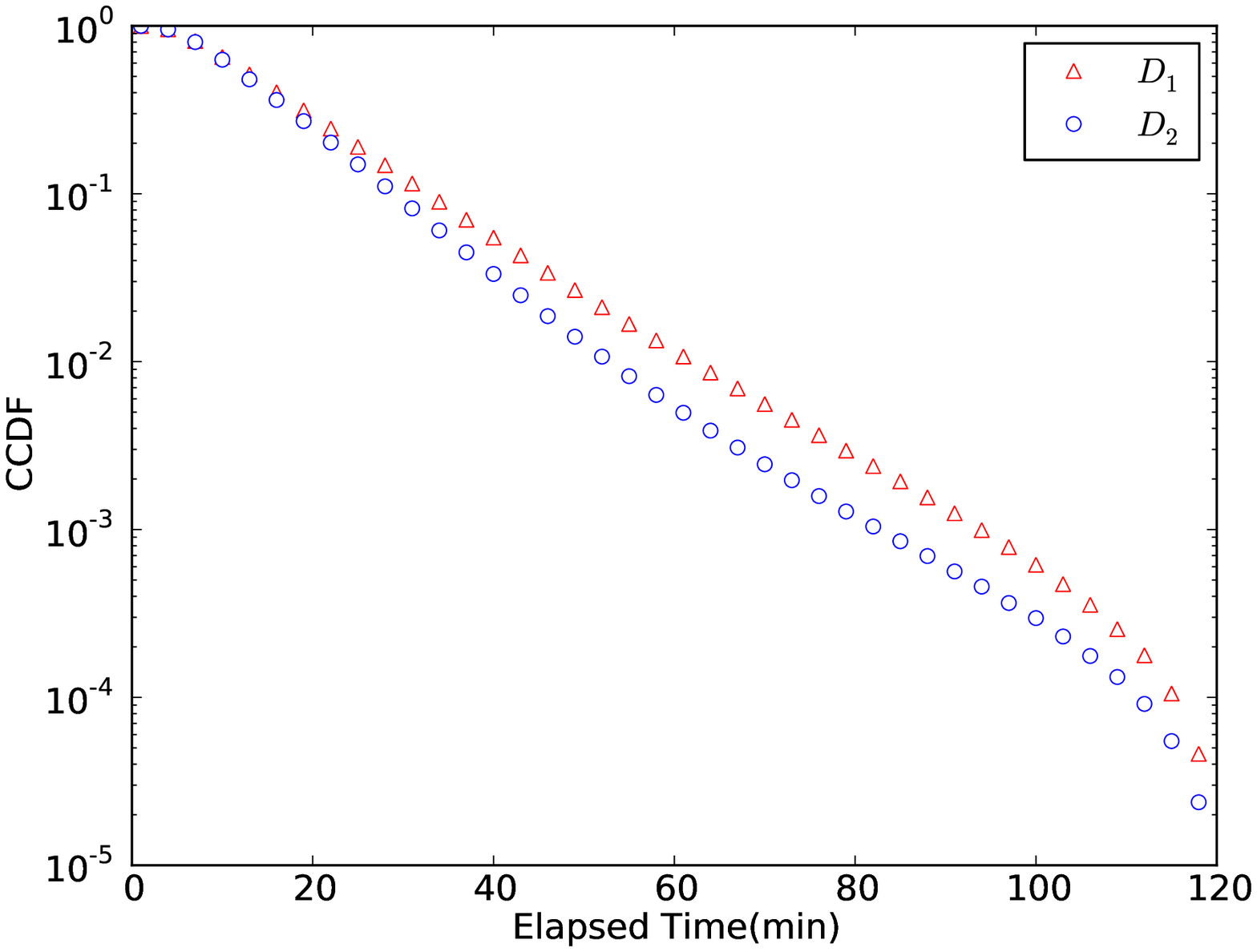}
  } \caption{ Elapsed time of trajectories. (a) The probability
    density function $P(\Delta T)$ obtained for studied dataset $D_1$
    in semi-log scale. The red dashed line indicates an exponential
    with the measured exponent $\lambda=0.0797$. The blue dotted line represents an exponential
    with measured exponent $\lambda = 0.0692$. (b) The CCDFs of elapsed time for both datasets.}
\end{figure}

Elapsed time $\Delta T$ means the time that passengers spend on travelling
from their origins to destinations. Given the origin and destination, the
elapsed time may be influenced by many factors such as current
transportation contexts, length of path chosen and habits of driving,
etc. We compute the elapsed time of trajectories in both
datasets. The distribution of elapsed time from the dataset $D_1$ is
shown in Figure \ref{fig:elapse_a}. Similar to displacements,
the distribution of elapsed time increases when $\Delta T \leq 7\text{ min}$,
and then decreases dramatically. There are about 98.9\% trajectories
in D1 and 99.5\% ones in D2 with elapsed time of less than 60
min. As we know, short elapsed time was mainly caused by most
trips with small displacements in urban areas, while long elapsed time
was often caused by traffic jams and long trips between urban areas and suburbs. 
Likewise, we partition the distribution into two parts according to the elapsed time of 60
min. In order to decide good fits, the
method of model selection mentioned in Section \ref{sec:pre} is used
as well. The detailed results are listed in Table \ref{tab:modelsel1},
illustrating that the first parts of distributions can be well approximated by
exponential fits in both datasets, while the latter parts of distributions in
$D_1$ and $D_2$ are inclined to an exponential and a power-law
respectively. Because we mainly focus on the movements in urban
areas, it can be concluded from the first parts of distributions that elapsed time of trips in urban areas
agrees with an exponential very well.

\begin{table}[htbp]
\begin{threeparttable}
  \caption{Results of model selection for elapsed time in both
    datasets.}
  \label{tab:modelsel1} 
  \centering 
  \begin{tabular}{ccccc}
   \hline
   {Part} & \multicolumn{2}{c}{Model}  & $D_1$ & $D_2$\\
   \hline
   \multirow{8}{*}{First part} & \multirow{4}{*}{Power law} & {MLE for $\alpha$} & {1.5549} & {1.7057}\\
   {} & {} & (95\% CI)\tnote{\S} & (1.5539, 1.5559) & (1.7046, 1.7069)\\
   {} & {} & $R^2$\ \tnote{\dag} & 0.8523 & 0.8318\\
   {} & {} & $W_{pow}$\tnote{\ddag} & 0 & 0\\
   \cline{2-5}
   {} & \multirow{4}{*}{Exponential} & {MLE for $\lambda$} & \textbf{0.0797} & \textbf{0.0912}\\
   {} & {} & (95\% CI) & (0.0796, 0.0797) & (0.0911, 0.0912)\\
   {} & {} & $R^2$ & 0.9927 & 0.9868\\
   {} & {} & $W_{exp}$\tnote{\ddag} & 1 & 1\\
   \hline
  \multirow{8}{*}{Latter part} & \multirow{4}{*}{Power law} & {MLE for $\alpha$} & {5.4924} & {\textbf{5.7775}}\\
   {} & {} & (95\% CI) & (5.4602, 5.5246) & (5.7243, 5.8306)\\
   {} & {} & $R^2$ & 0.9965 & 0.9991\\
   {} & {} & $W_{pow}$ & 0 & 1\\
   \cline{2-5}
   {} & \multirow{4}{*}{Exponential} & {MLE for $\lambda$} & \textbf{0.0692} & 0.0727\\
   {} & {} & (95\% CI) & (0.0688, 0.0696) & (0.0720, 0.0735)\\
   {} & {} & $R^2$ & 0.9984 & 0.9907\\
   {} & {} & $W_{exp}$ & 1 & 0\\
   \hline
\end{tabular} 
 \begin{tablenotes}
   \item[\S] A 95\% confidence interval.
   \item[\dag] Coefficient of determination to measure the goodness of fit of a model.
   \item[\ddag] Akaike weights representing relative likelihoods of models.
 \end{tablenotes}
\end{threeparttable}
\end{table}

\begin{figure}[htbp]
  \centering
  \includegraphics[scale=.4]{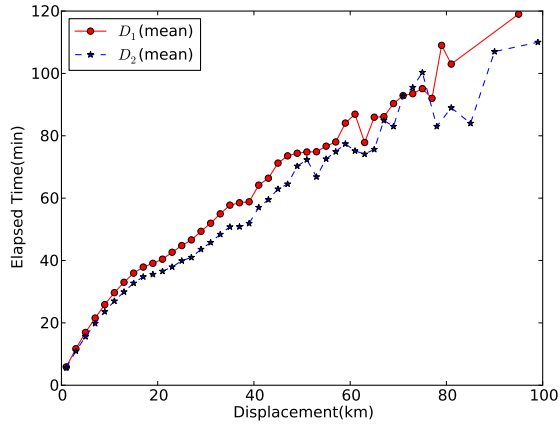}
  \caption{Comparison of traffic congestion status between the two datasets. The
    red/blue point represents the mean of elapsed time for certain 
    displacement.}
  \label{fig:disp_elapse_comp}
\end{figure}

Furthermore, in order to explore the dynamic trends of elapsed time,
we plot the distributions of elapsed time for both datasets
respectively. As shown in Figure \ref{fig:elapse_b}, it seems that
the distribution in $D_2$ drops more quickly than that in $D_1$, indicating that passengers have spent more time on taking taxis on
average. At the same time, the displacements people travelled have not changed too
much as mentioned in Subsection \ref{subsec:disp}. Besides, we can see from 
Figure \ref{fig:disp_elapse_comp} that the mean of elapsed time passengers spent on the same displacement in $D_1$ is basically longer than 
one in $D_2$. So it can be concluded that the
transportation conditions in Beijing have become worse since 2008,
which agrees with the facts very well. From Figure
\ref{fig:disp_elapse_comp}, we also notice that the rate of growth of
elapsed time becomes slower especially when the displacement is larger
than 20 km, which implies a higher average speed. It is reasonable
since trips with large displacements are expected to be away from traffic jams in urban areas.

\subsection{Correlation between displacement and elapsed time}
The first parts of distributions of displacement and elapsed time both follow
exponentials very well. In this subsection we will discuss the correlation
between them. Note that we use the dataset $D_1$ for experiments
below. The similar phenomena also apply to the dataset $D_2$.

\begin{figure}[htbp]
  \centering
  \includegraphics[scale=.4]{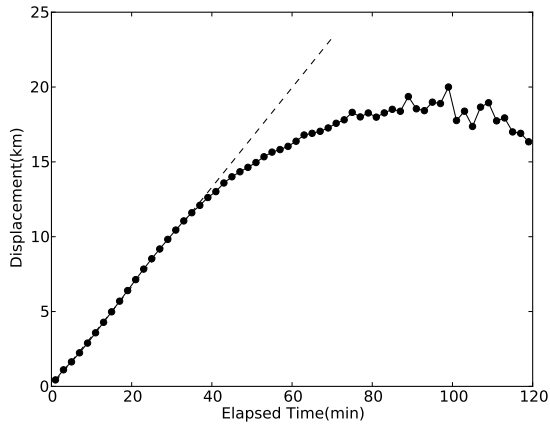}
  \caption{Correlation between displacement and elapsed time. The
    black point represents the mean of displacements for certain
    elapsed time. The dashed line denotes a fit $\Delta l =\mu \Delta T$ with
    $\mu = 0.3326$.}
  \label{fig:elapse_disp}
\end{figure}

For every individual movement, the displacement $\Delta l$ and the
elapsed time $\Delta T$ have been measured. Therefore, there are a lot
of trajectories with the same elapsed time $\Delta T$, which often
correspond to different displacement $\Delta l$. The correlation can be
shown in Figure \ref{fig:elapse_disp}. From the graph, we can observe that the mean of
displacements increases with elapsed time. The growth rate becomes
slower gradually when elapsed time is large. It
demonstrate that the long elapsed time was often caused by traffic
congestion leading to smaller average speed. Especially, when $\Delta T
\le 40\text{ min}$, the means can be well fitted by a linear function $\Delta l
=\mu \Delta T$ with $\mu=0.3326$. Here it must be emphasized that the
relation between $\Delta l$ and $\Delta T$ is numerical approximation
and does not hold in general. As we know from the above
observations, displacements of 98\% trajectories are less than
20 km and elapsed times of 95\% trajectories are less than 40
min. Hence the relationship between exponential exponents of the first
parts of both distributions can be approximately derived as follows.

Given the exponents $\lambda_{l}$, $\lambda_{T}$ for the first parts
of distributions respectively, the fit can be
represented by
\[E(\Delta l | \Delta T) = \mu \Delta T\]  
Hence, 
\begin{eqnarray*}
\sum_{\Delta T}{E(\Delta l | \Delta T)P(\Delta T)} & = & \mu \sum_{\Delta T}{ \Delta T P(\Delta T)}\\
E(\Delta l) & = & \mu E(\Delta T)
\end{eqnarray*}
Because the first parts of distributions of displacement and elapsed
time are relevant with most trajectories and decay
exponentially, we can obtain the expectations of both distributions
approximately in the forms
\[E(\Delta l) = \frac{1}{\lambda_{l}}, E(\Delta T) = \frac{1}{\lambda_{T}}\]
So the relationship is given by
\[\lambda_{T}=\lambda_{l}{\mu}\]
From the Table \ref{tab:modelsel} and \ref{tab:modelsel1}, we can
acquire $\lambda_l = 0.2329, \lambda_T = 0.0797$. Also $\lambda_T$ can
be caculated from relationship $0.2329 * 0.3326 \approx 0.0775$. The value is
very close to the MLE for $\lambda_T$.

\subsection{Interevent time}
After carrying passengers from origins to destinations, taxis begin to
wander or wait in the roads in order to seek new passengers. The
interevent time often means the time spent on waiting for new
customers. Intuitively, interevent time can be used to indicate degree of taxis'
busyness during certain period of time. In fact, short interevent
time often means that there are more demands for travelling statistically. Therefore, to a large extent,
it enables to reflect human travel demands indirectly.

\begin{figure*}[htbp]
  \centering \subfloat[]{
    \label{fig:inter_a}
    \includegraphics[scale=.3]{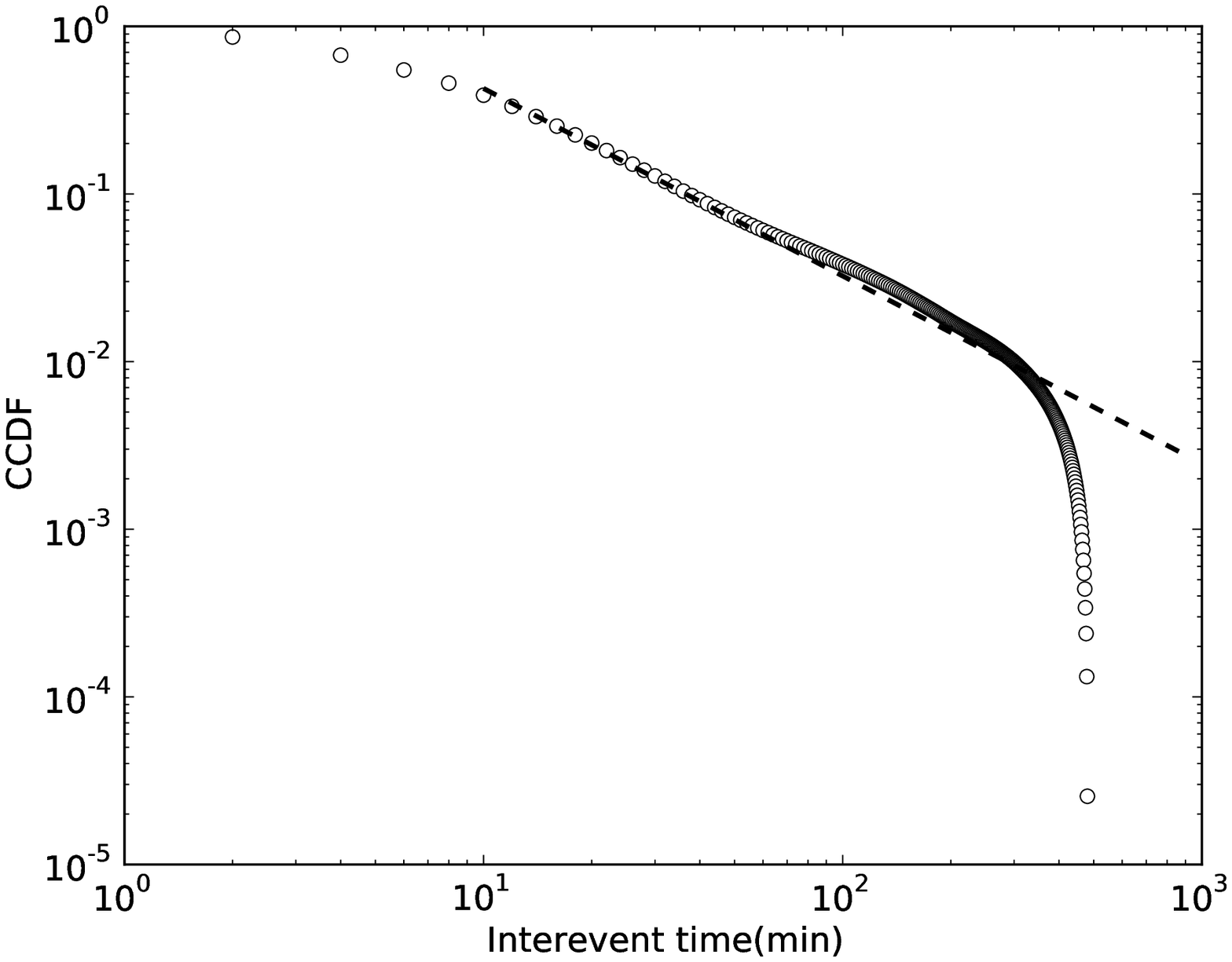}
  } \hspace{1pt} \subfloat[]{
    \label{fig:inter_b}
    \includegraphics[scale=.3]{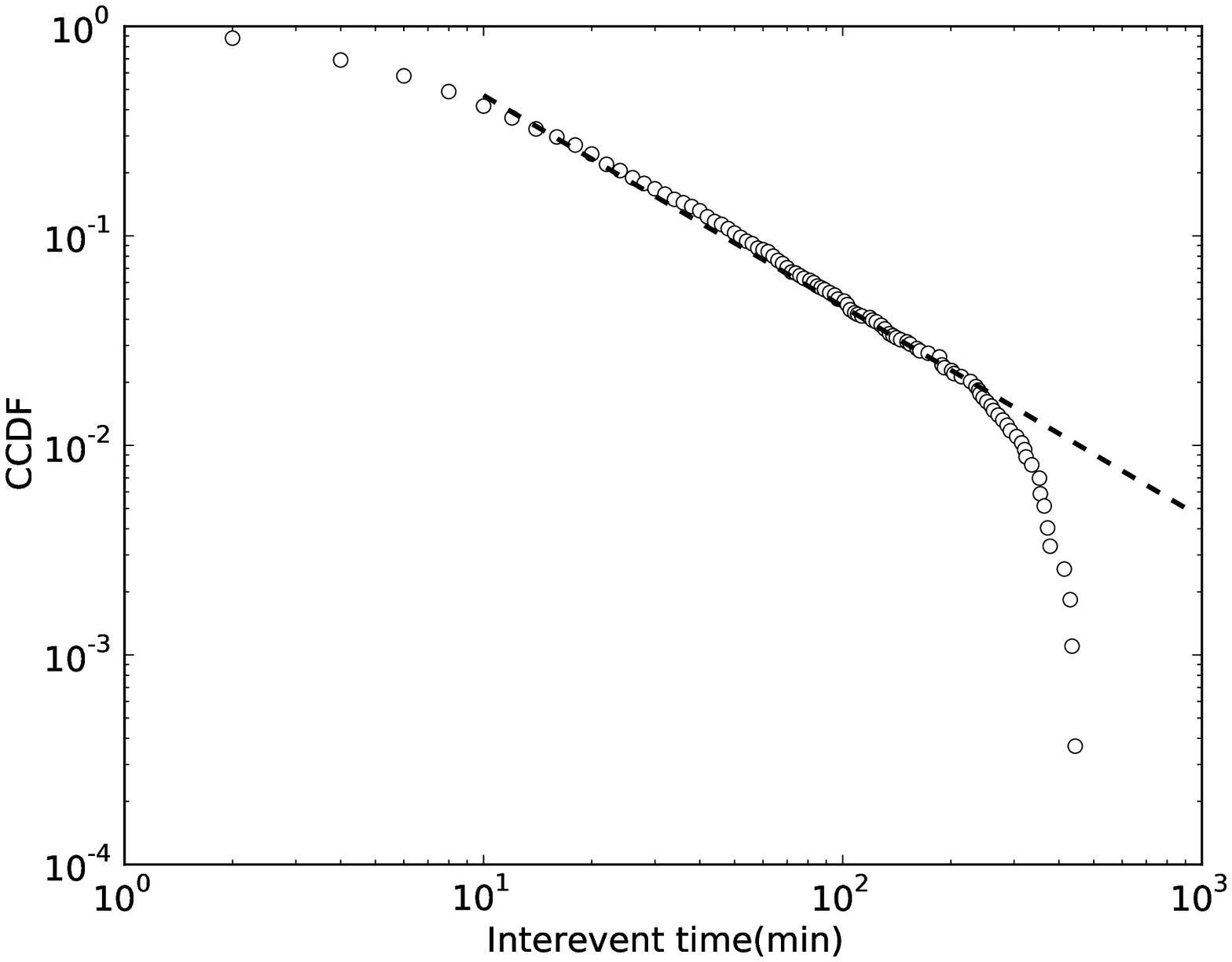}
  }\\ \subfloat[]{
    \label{fig:inter_c}
    \includegraphics[scale=.3]{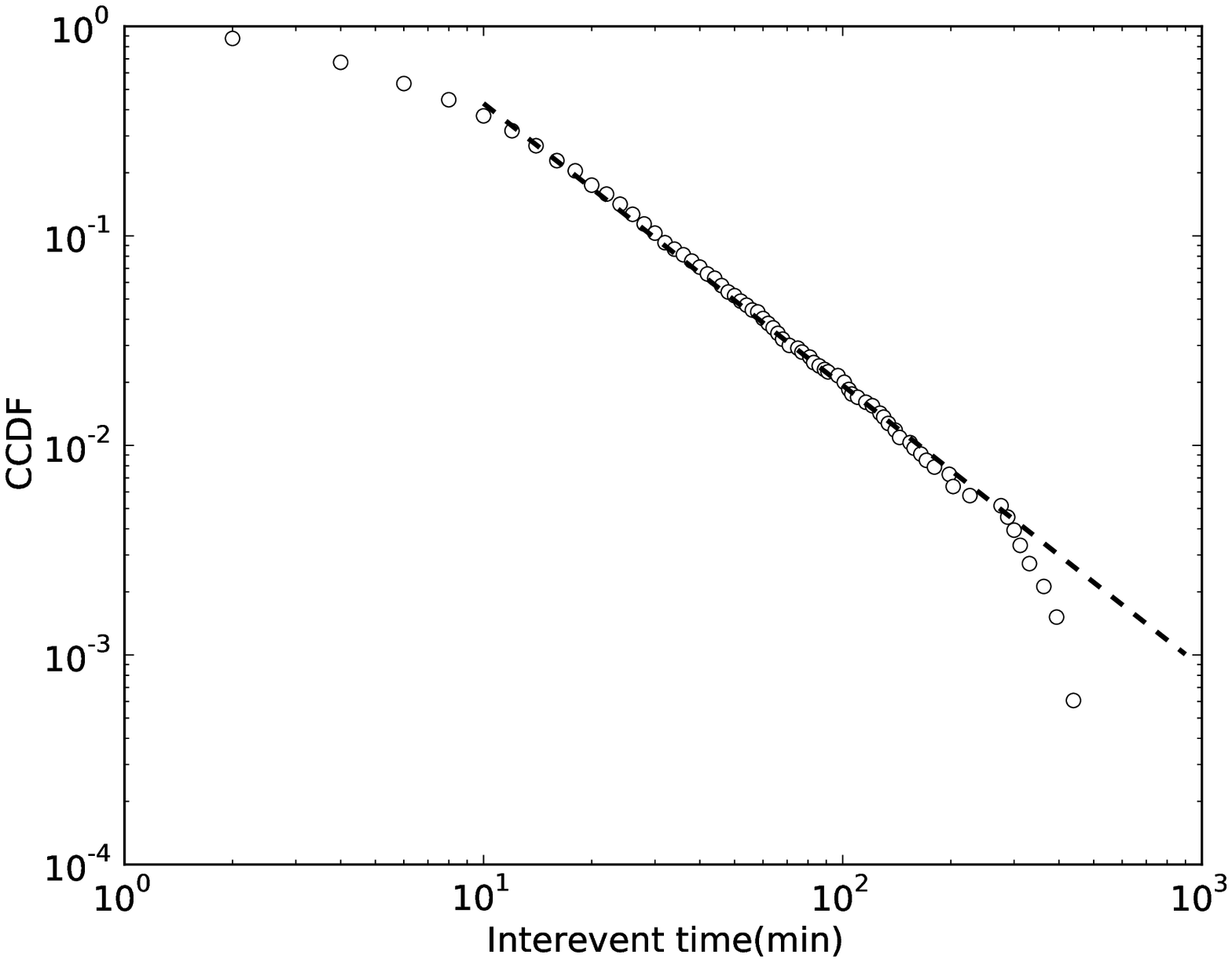}
  }\hspace{1pt} \subfloat[]{
    \label{fig:inter_d}
    \includegraphics[scale=.3]{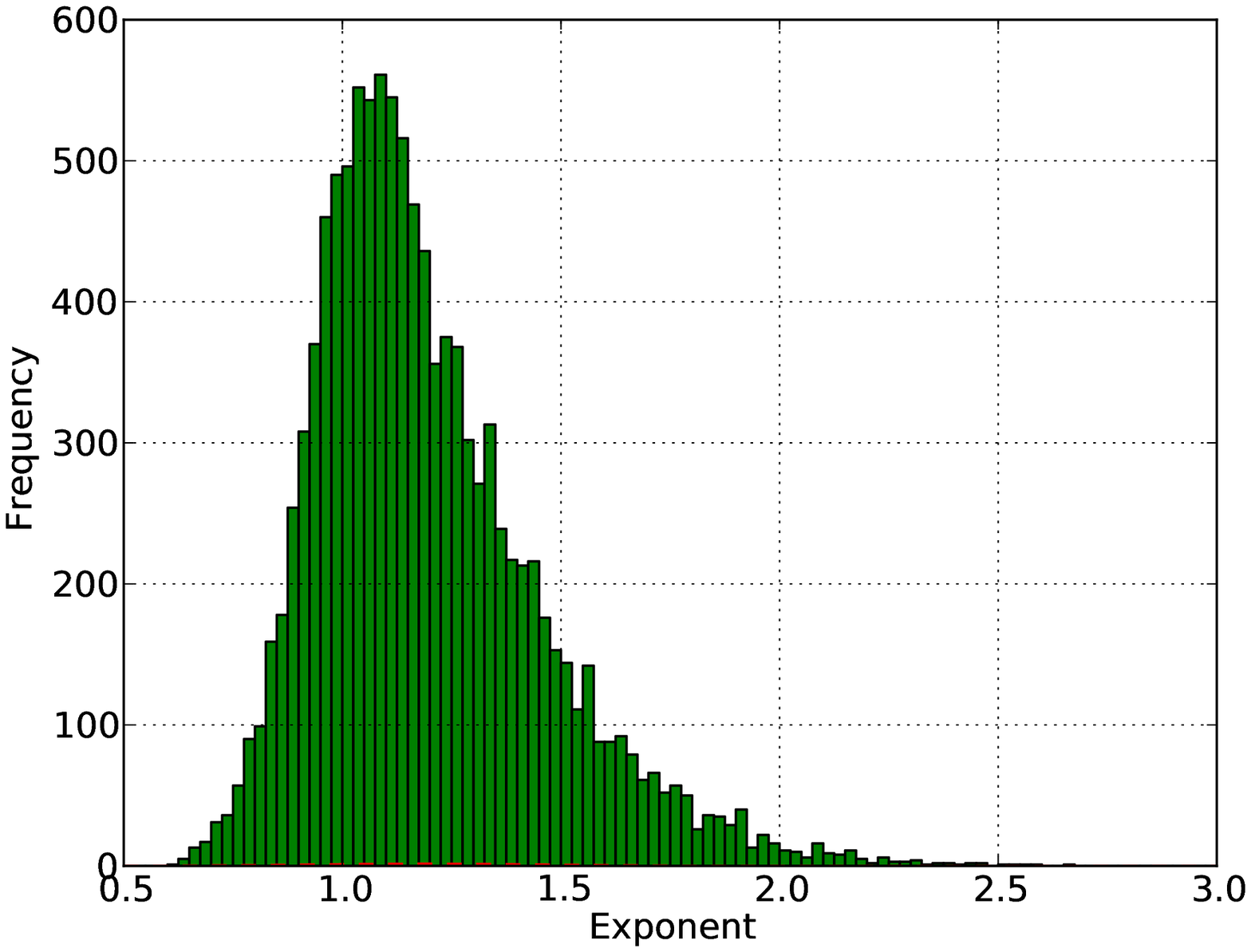}
  } \caption{Interevent time from the dataset
    $D_1$. (a) The CCDF for all taxis with exponent $\alpha =
    1.12$. (b-c) The CCDFs for two taxis with
    exponents $\alpha$ are 1.01 and 1.35 respectively. (d)
    The distribution of power-law exponents of CCDFs for all taxis with mean $\mu = 1.19$,
    standard deviation $\sigma = 0.26$.}
\end{figure*}

In order to characterize human travel demands, we compute the
interevent time $\tau$ from trajectories of $D_1$. The CCDF of
interevent time from dataset $D_1$ is shown in Figure 
\ref{fig:inter_a}. From the graph, we can see about 98\% of interevent
times are no more than 200 minutes. When interevent time is below
200 minutes, the curve fits a power-law very well. Then, it decreases
exponentially. The deviation from a power-law in the tail is caused by
two reasons. On the one hand, long interevent time often occurred
late at night when there were few travel demands. On the other, many
taxi drivers stopped working to have a rest at midnight.
Also, two different taxis are chosen randomly and
the CCDFs of interevent time are plotted in Figure
\ref{fig:inter_b} and \ref{fig:inter_c} respectively. Both curves
can be well approximated by power-laws and the two exponents are only slightly
different. Moreover, we plot the distribution of power-law exponents
of CCDFs for all taxis in Figure \ref{fig:inter_d}. As shown in
the figure, most taxis have similar power-law exponents around the
mean value 1.19. It must be remarked that the exponent obtained is
concerned with CCDF. So for probability density function (PDF), the exponent of
distribution is very close to 2.

In summary, it can be concluded that durations of taxis without 
passengers(inactivity) is close to be distributed according an inverse-square
power-law, while durations of taxis with passengers(activity) is well
approximated by an exponential. The results resemble ones discovered
in many animals with patterns of activity and inactivity where 
durations of inactivity follow inverse-square power-law distributions
approximately while durations of activity are fitted by exponential
distributions \cite{Reynolds2011}. There may be some subtle relations
between them deserving to be studied further.

\begin{figure}[htbp]
  \centering
  \includegraphics[scale=.4]{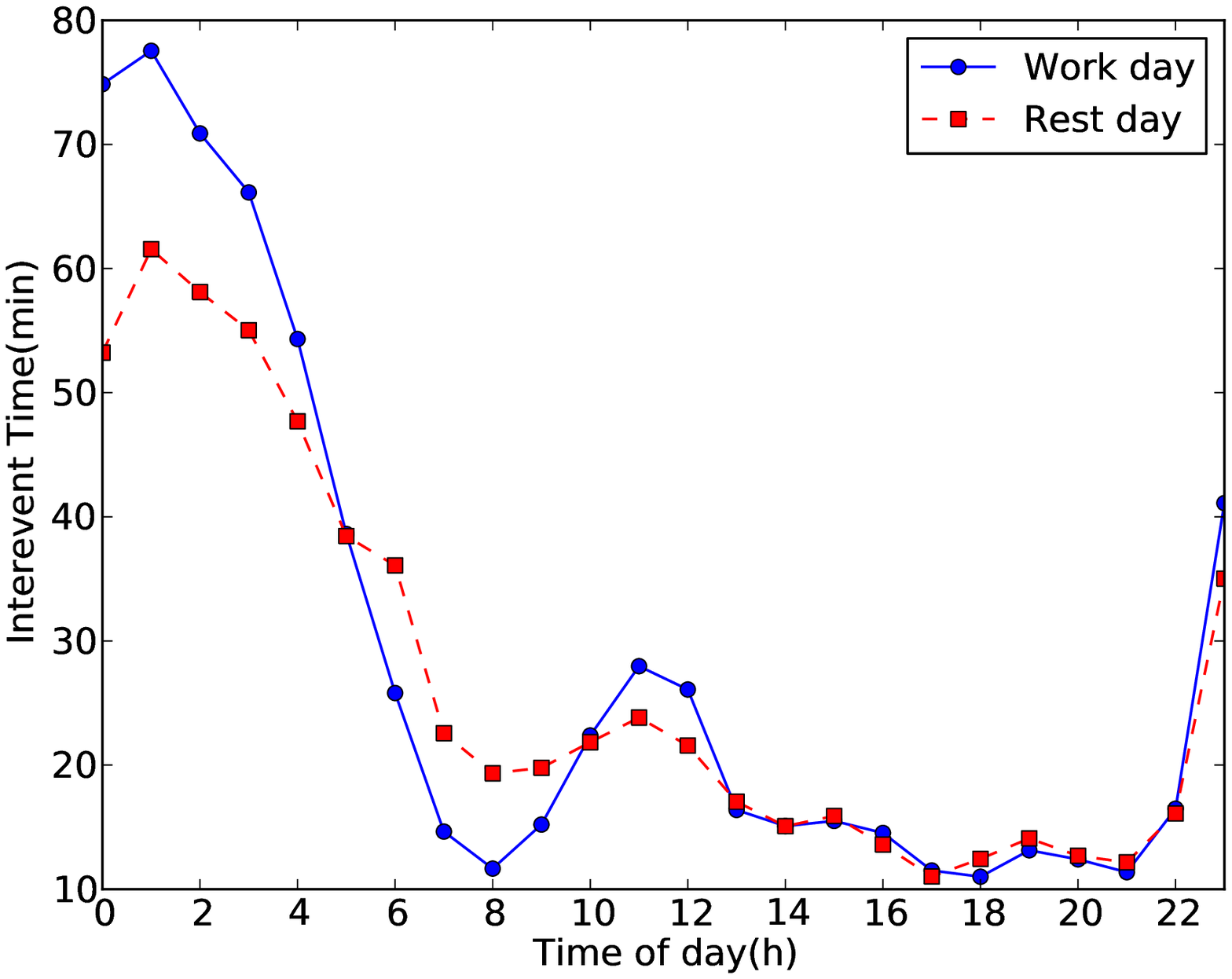}
  \caption{Mean value of interevent time for per hour in workdays and restdays.}
  \label{fig:wait_withhour}
\end{figure}

Furthermore, we divide the taxi trajectories into two categories based on
workday and rest day (including weekends and public holidays). Then we
calculate the mean values of interevent time occurred at different
hours, which are shown in Figure \ref{fig:wait_withhour}. As the graph
indicates, there are more than 30 minutes of interevent time during
the time slot of 23:00-5:00 in workday and rest day because people
often stay at home for a sleep and seldom have needs to travel. We
also notice that the interevent time during the time slot is shorter in rest day than in
workday, which is caused that people often readily go out
for entertainments in rest day. At the same time, people usually have
a rest for lunch resulting in longer interevent time during the time
slot of 11:00-12:00. As for workday, there are intensive trips occurred
in time slots of 7:00-9:00 and 17:00-19:00 corresponding the rush time when
people go to work and go off work. As for rest day, there are more
travel demands in the afternoon or evening than in the
morning. Therefore, it can be concluded that human travels in cities have
characteristic of bursts, behaving as a large amount of movements
emerge suddenly separated by long periods of inactivity because of strong periodic
variations of travel demands in cities.

The results of interevent time demonstrate that human travel demands
follow non-Poisson processes like many other human activities, for example, sending or
receiving Emails, web browsing, library loans \cite{Barabasi2005},
etc. In contrast with an exponential distribution, a heavy-tailed
distribution of interevent time allows for events occur frequently in
a relatively short period of time, then are inactive for a long
time. So the periodic variation of travel demands in
cities may be a part of reason to account for the results of interevent time.

\section{\label{sec:con}Conclusion and future work}
In this paper, we build models for 20 million trajectories collected from 10 thousand
taxis in urban areas in Beijing. The main contributions of this paper is threefold:
(i) models fitting the taxis' traveling displacements in urban areas tend to follow an exponential
distribution instead of a power-law, probably caused by the range of the movements
and economic effects; (ii) similarly, the elapsed time can
also be well approximated by an exponential distribution; (iii) bursty activities
are observed in modeling the interevent time between the trajectories, which also
help validate the bursty nature of human daily activities.

As future work, we are interested to further explore the two factors that affect
the trajectories of taxis' (i.e., the range of movements, and economic effects).
Besides, we observe that the quality of the mobility data heavily relies on length
of elapsed time for trajectories, and is dataset dependent. For example, elapsed
time for movement of banking notes are normally long, and trajectories of short
elapsed time indicate good data quality. To the opposite, elapsed time for trackings
of cell phone users' locations is usually short, and thus long elapsed time indicate
high quality of the trajectories. So we are interested to study to what extent the
limitations of the data influence the modeling of the distribution of displacements.
Last but not least, we would like to explore the relationship between mobility patterns
modeled from taxis trajectories, and mobility patterns observed in other datasets.

\section*{Acknowledgements}
We would like to thank the anonymous reviewers for their valuable suggestions.
We are also very grateful to Zhiyuan Cheng for helping us revise the manuscript.
The research was supported by the fund of the State Key Laboratory of
Software Development Environment (SKLSDE-2011ZX-02).





\bibliographystyle{model1-num-names}
\bibliography{humandyn}

\begin{thebibliography}{28}
\expandafter\ifx\csname natexlab\endcsname\relax\def\natexlab#1{#1}\fi
\providecommand{\bibinfo}[2]{#2}
\ifx\xfnm\relax \def\xfnm[#1]{\unskip,\space#1}\fi
\bibitem[{Rozenfeld et~al.(2008)Rozenfeld, Rybski, Andrade, Batty, Stanley, and
  Makse}]{Rozenfeld2008b}
\bibinfo{author}{H.~D. Rozenfeld}, \bibinfo{author}{D.~Rybski},
  \bibinfo{author}{J.~S. Andrade}, \bibinfo{author}{M.~Batty},
  \bibinfo{author}{H.~E. Stanley}, \bibinfo{author}{H.~A. Makse},
\newblock \bibinfo{title}{{Laws of population growth}},
\newblock \bibinfo{journal}{Proceedings of the National Academy of Sciences of
  the United States of America} \bibinfo{volume}{105} (\bibinfo{year}{2008})
  \bibinfo{pages}{18702--18707}.
\bibitem[{Jiang et~al.(2009)Jiang, Yin, and Zhao}]{Jiang2009}
\bibinfo{author}{B.~Jiang}, \bibinfo{author}{J.~Yin},
  \bibinfo{author}{S.~Zhao},
\newblock \bibinfo{title}{Characterizing the human mobility pattern in a large
  street network},
\newblock \bibinfo{journal}{Phys. Rev. E} \bibinfo{volume}{80}
  (\bibinfo{year}{2009}) \bibinfo{pages}{021136}.
\bibitem[{Agliari et~al.(2010)Agliari, Burioni, and Cassi}]{Agliari2010}
\bibinfo{author}{E.~Agliari}, \bibinfo{author}{R.~Burioni},
  \bibinfo{author}{D.~Cassi},
\newblock \bibinfo{title}{Word-of-mouth and dynamical inhomogeneous markets: an
  efficiency measure and optimal sampling policies for the pre-launch stage},
\newblock \bibinfo{journal}{IMA J. Manage. Math.} \bibinfo{volume}{21}
  (\bibinfo{year}{2010}) \bibinfo{pages}{67--83}.
\bibitem[{Hufnagel et~al.(2004)Hufnagel, Brockmann, and Geisel}]{Hufnagel2004}
\bibinfo{author}{L.~Hufnagel}, \bibinfo{author}{D.~Brockmann},
  \bibinfo{author}{T.~Geisel},
\newblock \bibinfo{title}{{Forecast and control of epidemics in a globalized
  world}},
\newblock \bibinfo{journal}{Proceedings of the National Academy of Sciences of
  the United States of America} \bibinfo{volume}{101} (\bibinfo{year}{2004})
  \bibinfo{pages}{15124--15129}.
\bibitem[{Lee et~al.(2009)Lee, Hong, Kim, Rhee, and Chong}]{Lee2009}
\bibinfo{author}{K.~Lee}, \bibinfo{author}{S.~Hong}, \bibinfo{author}{S.~J.
  Kim}, \bibinfo{author}{I.~Rhee}, \bibinfo{author}{S.~Chong},
\newblock \bibinfo{title}{{SLAW: A mobility model for human walks}},
\newblock in: \bibinfo{booktitle}{INFOCOM 2009}, \bibinfo{publisher}{IEEE},
  \bibinfo{year}{2009}, pp. \bibinfo{pages}{855--863}.
\bibitem[{Gonz\'{a}lez et~al.(2008)Gonz\'{a}lez, Hidalgo, and
  Barab\'{a}si}]{Gonzalez2008}
\bibinfo{author}{M.~Gonz\'{a}lez}, \bibinfo{author}{C.~Hidalgo},
  \bibinfo{author}{A.-L. Barab\'{a}si},
\newblock \bibinfo{title}{{Understanding individual human mobility patterns}},
\newblock \bibinfo{journal}{Nature} \bibinfo{volume}{453}
  (\bibinfo{year}{2008}) \bibinfo{pages}{779--782}.
\bibitem[{Song et~al.(2010)Song, Qu, Blumm, and Barab\'{a}si}]{Song2010}
\bibinfo{author}{C.~Song}, \bibinfo{author}{Z.~Qu}, \bibinfo{author}{N.~Blumm},
  \bibinfo{author}{A.-L. Barab\'{a}si},
\newblock \bibinfo{title}{{Limits of predictability in human mobility}},
\newblock \bibinfo{journal}{Science} \bibinfo{volume}{327}
  (\bibinfo{year}{2010}) \bibinfo{pages}{1018--1021}.
\bibitem[{Choujaa and Dulay(2010)}]{Choujaa2010}
\bibinfo{author}{D.~Choujaa}, \bibinfo{author}{N.~Dulay},
\newblock \bibinfo{title}{{Predicting human behaviour from selected mobile
  phone data points}},
\newblock in: \bibinfo{booktitle}{Proceedings of the 12th ACM international
  conference on Ubiquitous computing}, Ubicomp '10, \bibinfo{publisher}{ACM},
  \bibinfo{address}{Copenhagen, Denmark}, \bibinfo{year}{2010}, pp.
  \bibinfo{pages}{105--108}.
\bibitem[{Zheng et~al.(2008)Zheng, Li, Chen, Xie, and Ma}]{Zheng2008}
\bibinfo{author}{Y.~Zheng}, \bibinfo{author}{Q.~Li}, \bibinfo{author}{Y.~Chen},
  \bibinfo{author}{X.~Xie}, \bibinfo{author}{W.-Y. Ma},
\newblock \bibinfo{title}{{Understanding mobility based on GPS data}},
\newblock in: \bibinfo{booktitle}{Proceedings of the 10th international
  conference on Ubiquitous computing}, UbiComp '08, \bibinfo{publisher}{ACM},
  \bibinfo{address}{Seoul, Korea}, \bibinfo{year}{2008}, pp.
  \bibinfo{pages}{312--321}.
\bibitem[{Rhee et~al.(2008)Rhee, Shin, Hong, Lee, and Chong}]{Rhee2008}
\bibinfo{author}{I.~Rhee}, \bibinfo{author}{M.~Shin},
  \bibinfo{author}{S.~Hong}, \bibinfo{author}{K.~Lee},
  \bibinfo{author}{S.~Chong},
\newblock \bibinfo{title}{{On the Levy-walk nature of human mobility}},
\newblock in: \bibinfo{booktitle}{INFOCOM 2008}, \bibinfo{publisher}{IEEE},
  \bibinfo{year}{2008}, pp. \bibinfo{pages}{924--932}.
\bibitem[{Bazzani et~al.(2010)Bazzani, Giorgini, Rambaldi, Gallotti, and
  Giovannini}]{Bazzani2010}
\bibinfo{author}{A.~Bazzani}, \bibinfo{author}{B.~Giorgini},
  \bibinfo{author}{S.~Rambaldi}, \bibinfo{author}{R.~Gallotti},
  \bibinfo{author}{L.~Giovannini},
\newblock \bibinfo{title}{{Statistical laws in urban mobility from microscopic
  GPS data in the area of Florence}},
\newblock \bibinfo{journal}{Journal of Statistical Mechanics: Theory and
  Experiment} \bibinfo{volume}{2010} (\bibinfo{year}{2010})
  \bibinfo{pages}{P05001}.
\bibitem[{Jiang and Jia(2011)}]{Jiang2011a}
\bibinfo{author}{B.~Jiang}, \bibinfo{author}{T.~Jia}, \bibinfo{title}{Exploring
  human mobility patterns based on location information of us flights},
  \bibinfo{year}{2011}. \bibinfo{note}{ArXiv:1104.4578v2 [physics.data-an]}.
\bibitem[{Veloso et~al.(2011)Veloso, Phithakkitnukoon, Bento, Fonseca, and
  Olivier}]{Veloso2011}
\bibinfo{author}{M.~Veloso}, \bibinfo{author}{S.~Phithakkitnukoon},
  \bibinfo{author}{C.~Bento}, \bibinfo{author}{N.~Fonseca},
  \bibinfo{author}{P.~Olivier},
\newblock \bibinfo{title}{{Exploratory Study of Urban Flow using Taxi Traces}},
\newblock in: \bibinfo{booktitle}{The First Workshop on Pervasive Urban
  Applications (PURBA) 2011}.
\bibitem[{Kim et~al.(2006)Kim, Kotz, and Kim}]{Kim2006}
\bibinfo{author}{M.~Kim}, \bibinfo{author}{D.~Kotz}, \bibinfo{author}{S.~Kim},
\newblock \bibinfo{title}{{Extracting a mobility model from real user traces}},
\newblock in: \bibinfo{booktitle}{INFOCOM 2006}, \bibinfo{publisher}{IEEE},
  \bibinfo{year}{2006}, pp. \bibinfo{pages}{1--13}.
\bibitem[{Cheng et~al.(2011)Cheng, Caverlee, and Lee}]{Cheng2011}
\bibinfo{author}{Z.~Cheng}, \bibinfo{author}{J.~Caverlee},
  \bibinfo{author}{K.~Lee},
\newblock \bibinfo{title}{{Exploring Millions of Footprints in Location Sharing
  Services}},
\newblock in: \bibinfo{booktitle}{ICWSM 2011}, \bibinfo{number}{Cholera}.
\bibitem[{Cho et~al.(2011)Cho, Myers, and Leskovec}]{Cho2011}
\bibinfo{author}{E.~Cho}, \bibinfo{author}{S.~Myers},
  \bibinfo{author}{J.~Leskovec},
\newblock \bibinfo{title}{{Friendship and Mobility: User Movement In
  Location-Based Social Networks}},
\newblock in: \bibinfo{booktitle}{KDD 2011}.
\bibitem[{Brockmann et~al.(2006)Brockmann, Hufnagel, and
  Geisel}]{Brockmann2006}
\bibinfo{author}{D.~Brockmann}, \bibinfo{author}{L.~Hufnagel},
  \bibinfo{author}{T.~Geisel},
\newblock \bibinfo{title}{{The scaling laws of human travel}},
\newblock \bibinfo{journal}{Nature} \bibinfo{volume}{439}
  (\bibinfo{year}{2006}) \bibinfo{pages}{462--465}.
\bibitem[{Viswanathan et~al.(1996)Viswanathan, Afanasyev, Buldyrev, Murphy,
  Prince, and Stanley}]{Viswanathan1996}
\bibinfo{author}{G.~Viswanathan}, \bibinfo{author}{V.~Afanasyev},
  \bibinfo{author}{S.~Buldyrev}, \bibinfo{author}{E.~Murphy},
  \bibinfo{author}{P.~Prince}, \bibinfo{author}{H.~E. Stanley},
\newblock \bibinfo{title}{{Levy flight search patterns of wandering
  albatrosses}},
\newblock \bibinfo{journal}{Nature} \bibinfo{volume}{381}
  (\bibinfo{year}{1996}) \bibinfo{pages}{413--415}.
\bibitem[{Atkinson et~al.(2002)Atkinson, Rhodes, Macdonald, and
  Anderson}]{Atkinson2002}
\bibinfo{author}{R.~P.~D. Atkinson}, \bibinfo{author}{C.~J. Rhodes},
  \bibinfo{author}{D.~W. Macdonald}, \bibinfo{author}{R.~M. Anderson},
\newblock \bibinfo{title}{{Scale-free dynamics in the movement patterns of
  jackals}},
\newblock \bibinfo{journal}{Oikos} \bibinfo{volume}{98} (\bibinfo{year}{2002})
  \bibinfo{pages}{134--140}.
\bibitem[{Song et~al.(2010)Song, Koren, Wang, and Barab\'{a}si}]{Song2010a}
\bibinfo{author}{C.~Song}, \bibinfo{author}{T.~Koren},
  \bibinfo{author}{P.~Wang}, \bibinfo{author}{A.-L. Barab\'{a}si},
\newblock \bibinfo{title}{{Modelling the scaling properties of human
  mobility}},
\newblock \bibinfo{journal}{Nature Physics} \bibinfo{volume}{6}
  (\bibinfo{year}{2010}) \bibinfo{pages}{818--823}.
\bibitem[{Han et~al.(2011)Han, Hao, Wang, and Tao}]{Han2011}
\bibinfo{author}{X.~Han}, \bibinfo{author}{Q.~Hao}, \bibinfo{author}{B.~Wang},
  \bibinfo{author}{Z.~Tao},
\newblock \bibinfo{title}{{Origin of the scaling law in human mobility:
  Hierarchy of traffic systems}},
\newblock \bibinfo{journal}{Physical Review E} \bibinfo{volume}{83}
  (\bibinfo{year}{2011}) \bibinfo{pages}{2--6}.
\bibitem[{Johnson and Omland(2004)}]{Johnson2004}
\bibinfo{author}{J.~B. Johnson}, \bibinfo{author}{K.~S. Omland},
\newblock \bibinfo{title}{{Model selection in ecology and evolution}},
\newblock \bibinfo{journal}{Trends in Ecology \& Evolution}
  \bibinfo{volume}{19} (\bibinfo{year}{2004}) \bibinfo{pages}{101--108}.
\bibitem[{Hastie et~al.(2008)Hastie, Tibshirani, and Friedman}]{Hastie2008}
\bibinfo{author}{T.~Hastie}, \bibinfo{author}{R.~Tibshirani},
  \bibinfo{author}{J.~Friedman}, \bibinfo{title}{{The elements of statistical
  learning}}, \bibinfo{publisher}{Springer}, \bibinfo{year}{2008}.
\bibitem[{Clauset and {Shalizi, Cosma Rohilla Newman}(2009)}]{Clauset2009}
\bibinfo{author}{A.~Clauset}, \bibinfo{author}{M.~E.~J. {Shalizi, Cosma Rohilla
  Newman}},
\newblock \bibinfo{title}{{Power-law distributions in empirical data}},
\newblock \bibinfo{journal}{SIAM Rev.} \bibinfo{volume}{51}
  (\bibinfo{year}{2009}) \bibinfo{pages}{661--703}.
\bibitem[{Edwards et~al.(2007)Edwards, Phillips, Watkins, Freeman, Murphy,
  Afanasyev, Buldyrev, da~Luz, Raposo, Stanley, and Viswanathan}]{Edwards2007}
\bibinfo{author}{A.~M. Edwards}, \bibinfo{author}{R.~A. Phillips},
  \bibinfo{author}{N.~W. Watkins}, \bibinfo{author}{M.~P. Freeman},
  \bibinfo{author}{E.~J. Murphy}, \bibinfo{author}{V.~Afanasyev},
  \bibinfo{author}{S.~V. Buldyrev}, \bibinfo{author}{M.~G.~E. da~Luz},
  \bibinfo{author}{E.~P. Raposo}, \bibinfo{author}{H.~E. Stanley},
  \bibinfo{author}{G.~M. Viswanathan},
\newblock \bibinfo{title}{{Revisiting L\'{e}vy flight search patterns of
  wandering albatrosses, bumblebees and deer}},
\newblock \bibinfo{journal}{Nature} \bibinfo{volume}{449}
  (\bibinfo{year}{2007}) \bibinfo{pages}{1044--1048}.
\bibitem[{Mashanova et~al.(2010)Mashanova, Oliver, and Jansen}]{Mashanova2010}
\bibinfo{author}{A.~Mashanova}, \bibinfo{author}{T.~H. Oliver},
  \bibinfo{author}{V.~A.~A. Jansen},
\newblock \bibinfo{title}{Evidence for intermittency and a truncated power law
  from highly resolved aphid movement data},
\newblock \bibinfo{journal}{J. R. Soc. Interface} \bibinfo{volume}{7}
  (\bibinfo{year}{2010}) \bibinfo{pages}{199--208}.
\bibitem[{Reynolds(2011)}]{Reynolds2011}
\bibinfo{author}{A.~Reynolds},
\newblock \bibinfo{title}{On the origin of bursts and heavy tails in animal
  dynamics},
\newblock \bibinfo{journal}{Physica A: Statistical Mechanics and its
  Applications} \bibinfo{volume}{390} (\bibinfo{year}{2011})
  \bibinfo{pages}{245--249}.
\bibitem[{Barab\'{a}si(2005)}]{Barabasi2005}
\bibinfo{author}{A.-L. Barab\'{a}si},
\newblock \bibinfo{title}{{The origin of bursts and heavy tails in human
  dynamics}},
\newblock \bibinfo{journal}{Nature} \bibinfo{volume}{435}
  (\bibinfo{year}{2005}) \bibinfo{pages}{207--211}.

\end{thebibliography}







\end{document}